\documentclass[amsmath,amssymb, aps]{revtex4-2} 
\setcitestyle{square}
\usepackage{url}
\usepackage{hyperref}
\usepackage{lineno}
\usepackage{microtype}
\usepackage{subcaption}
\usepackage[onehalfspacing]{setspace}
\usepackage{chemist}
\usepackage{chemformula}
\usepackage{float}
\usepackage{caption}
\begin{document}

\title {Mapping dissolved carbon in space and time: An experimental technique for the measurement of pH and total carbon concentration in density driven convection of CO$_2$ dissolved in water} 
\author{Hilmar Yngvi Birggison$^{1}$}
\author{Yao Xu$^{1}$}
\email{yaox@fys.uio.no}
\author{Marcel Moura$^{1}$} 
\author{Eirik Grude Flekkøy$^{1,2}$}
\author{Knut J{\o}rgen  M{\aa}l{\o}y$^{1,3}$}
\affiliation{$^1$PoreLab, Department of Physics, The Njord Centre, University of Oslo, Oslo Norway}
\affiliation{$^2$PoreLab, Department of Chemistry, Norwegian University of Science and Technology, Trondheim, Norway
}
\affiliation{$^3$PoreLab, Department of Geoscience and Petroleum, Norwegian University of Science and Technology, Trondheim, Norway}

\begin{abstract}
We present an experimental technique for determining the  pH and the total carbon concentration when \ch{CO2} diffuses and flows in water.  
The technique employs three different pH indicators, which, when combined with an image analysis technique, provides a dynamic range in pH from 4.0 to 9.5. In contrast to usual techniques in which a single pH indicator is used, the methodology presented allows not only to produce a binary classification (pH larger or smaller than a given threshold) but to access a much more complete continuous spatial distribution of pH and concentration levels in the system. We calibrate the method against benchmark solutions and further demonstrate its potential by measuring the pH and total carbon concentration in a density driven convection (DDC) of carbon-enriched water. The motivation for testing the method in this particular experiment comes from the fact that DDC plays a pivotal role in the efficiency of engineered carbon storage processes. The application of the technique presented here provided a direct window for the analysis of the spatial distribution of captured carbon in the DDC flow.

\end{abstract}

\maketitle

\date{\today}

\section{Introduction and motivation}

With the current state  of our energy affairs, and the climate risk associated with greenhouse gas emissions, scientists keep pursuing different ways of pushing our energy society towards a greener future with lower emissions. In tandem with a goal for reducing emissions is the relatively new approach of carbon capture and storage (CCS), which is an approach that can help achieve carbon neutrality, and although unfeasible today, could in theory be carbon negative.

One of the most important considerations of CCS is how the CO$_2$ is actually trapped for storage, as not to leak and contaminate the surrounding environment with carbonic acid. In the literature, the process is often broken down into several trapping mechanics \cite{snabjornsdottir2020,metz2005}. Initially, CO$_2$ is physically trapped by being injected below a caprock of low permeability. As the name suggests, this implies the injection site has a suitable geometry and low enough permeability to sufficiently contain the buoyant CO$_2$ from rising upwards. These traps commonly consist of folded or fractured rock.

When the injected CO$_2$ comes in contact with formation water, solubility trapping will also contribute to the carbon storage process. Solubility trapping refers to trapping by dissolution into liquids already present in the geological feature. The precise details of this depend heavily on the conditions \cite{king2005}. For example higher pressures and supercritical conditions speed up the dissolution process. One clear desirable aspect of solubility trapping is that the CO$_2$ enters the water phase, and dissolved CO$_2$ is thus not prone to seeping out of the reservoir due to buoyancy \cite{szulczewski2012,metz2005}. On the contrary, solubility trapping increases fluid density and the affected liquid will instead sink, giving rise to the phenomenon termed Density Driven Convection (DDC). This flow mechanism is key for enhancing the efficiency of solubility trapping, since it can accelerate the diffusive mixing that drives the dissolution process.

Consider a system consisting of gaseous Carbon Dioxide (CO$_2$) and liquid water. At ambient conditions a chemical equilibrium exists which causes the CO$_2$ to dissolve and diffuse into the water and form carbonic acid. The rate of dissolution depends on the CO$_2$ concentration at the gas-liquid interface and thus slows down and stagnates when the equilibrium is reached. However when one couples these dissolution dynamics to a flow field, something interesting happens. Since carbonic acid slightly increases the density of the liquid, the chemical dissolution at the boundary will cause a natural convective flow - which in turn brings more ”fresh” liquid to the boundary. We call this phenomena Density Driven Convection (DDC) \cite{szulczewski2012,dewit2016,thomas2015,riaz2006,backhaus2011,macminn2013,wooding1969,faisal2013,faisal2015,erfani2022,hidalgo2013}. This has the effect of increasing the potential rate of CO$_2$ uptake of the system compared to the purely diffusive case.
Understanding these dynamics better is paramount to understanding the hydrodynamic effects involved with carbon capture and storage (CCS), since CCS is often realized by pumping CO$_2$ into underground  formations sealed by some cap rock. Because of buoyancy the gas will seek to the top and the previously described dynamics take place at the gas-liquid interface and result in an instability with plume patterns \cite{szulczewski2012,dewit2016,thomas2015,riaz2006,backhaus2011,macminn2013,wooding1969,faisal2013,faisal2015,erfani2022,hidalgo2013}.
Much of the experimental work on plume patterns in density driven convection of CO$_2$ dissolved in water relies on the use of pH color indicators to visualize the convection plume patterns  in 2D Hele-Shaw cells \cite{kneafsey2010,outeda2014,faisal2013,mojtaba2014,thomas2015,faisal2015,faisal2013}. These studies can relatively easily show the CO$_2$ affected fluid regions, and how they propagate in time. This offers the possibility of tracking plume locations, lengths, areas etc. Compared to their relative simplicity, these experiments can give much intuitive insight into the complex plume patterns observed during DDC. However, questions have been raised about the methodology of the visualization techniques employed in some studies. Thomas et al. demonstrated that the resulting morphology of the convection plumes largely depends on the color indicator being used \cite{thomas2015}. This was determined by repeated experiments in Hele-Shaw cells colored by Bromocresol Green and Bromocresol Purple. Additionally, the authors compared the colored plume morphology obtained via a color indicator to fluctuations in refractive index obtained by schlieren imaging \cite{settles2001}.  Among the conclusions were that the choice of indicator does in fact dictate what the imaged patterns will look like, but did not appear to have an effect on the dynamics themselves.  This comes from the fact that the previous color indicator approaches have no meaningful way of estimating an amount of dissolved carbon, and as such, the collected data more resembles a binary image of an affected vs an unaffected area. This is our core motivation: to try and  improve on the color indicator approach, in search of data more suitable to quantitative analysis. 

Recent investigations have been directed toward achieving quantitative measurements of pH and carbon concentration. Research has explored refractive index matching and planar-laser-induced fluorescence to detect the pH \cite{brouzet2022co}. However, the quantitative pH determination was still difficult because the pH values strongly depended on the applied fluorescence functions. Other research has tried to map the pH values by the solution's color from Bromocresol Purple. It built a calibration color scale by the  [red(R) green(G) blue(B)] values of a set of aqueous solutions with known pH \cite{de2021bi}. The calibration curve was subsequently converted from the color maps into the pH maps.  In addition, the studies also endeavored to correlate pH with carbon concentration\cite{brouzet2022co,de2021bi}. However, performing the quantitative assessment of the concentration profile is arduous. As a result, tests were carried out to determine total dissolved \ch{CO2} by detecting pressure changes in the \ch{CO2} gas phase. This method only yields an overall dissolution, not a spatial or dynamic \ch{CO2} concentration
\cite{mojtaba2014,brouzet2022co,taheri2021effect,tang2019experimental}.

In this paper, we describe an experimental technique for measuring the pH and total carbon concentration in water and apply the technique to measure the carbon concentration in a DDC flow cell.
We begin with an explanation in Section \ref{section2} of the basic chemistry of the equilibrium reaction between CO$_2$ and water and the relationship between pH and total carbon concentration, assuming that the various forms of dissolved carbon are in local equilibrium.
Section \ref{section3} describes the experimental model system and the technique for capturing the local pH concentration and total carbon concentration using three different pH indicators and image analysis. In Section \ref{section4}, we apply the technique to the study of Density Driven Convection in a Hele-Shaw cell. We demonstrate how our methodology can be used to measure the pH and total carbon concentration in the sinking plumes of carbon-rich water. We finally produce a spatial map of dissolved carbon concentration for the experiment. 

\section{A brief chemistry interlude: connecting acidity and total carbon concentration} \label{section2}
As a first step in the development of our technique, one must understand the basic chemistry of the equilibrium reaction between CO$_2$ and water which is an example of a more involved system of equilibrium reactions \cite{henry1803}. This equilibrium is of interest to various fields of the natural sciences, as it dictates processes ranging from how oceans acidify due to increased carbon emissions to how living organisms regulate their pH by breathing.
Consider a body of water in contact with the atmosphere. Assume the idealized glass of water is initially pure, containing no CO$_2$ or other dissolved species. Conceptually one can think of the acidifying process as the following reactions. \\ \\
Gas  dissolution: 

\begin{equation}
CO_{2 (g)}\rightleftarrows CO_{2(aq)} 
\end{equation}
Hydration:
\begin{equation}
CO_{2(aq)} + H_2O   \rightleftarrows H_2CO_{3(aq)}
\end{equation}
First dissociation:
\begin{equation}
H_2CO_{3(aq)} \rightleftarrows H^+ + HCO_{3(aq)}^{-}
\end{equation}
Second dissociation:

\begin{equation}
HCO_{3(aq)}^{-} \rightleftarrows H^+ +CO_{3 (aq)}^{2-}
\end{equation}

The four aforementioned chemical equilibria describe how gaseous CO$_2$ can interact with water solutions to acidify them. The  equilibrium constants for the corresponding reaction and the corresponding equilibrium equations are shown in Tab.\ref{table:1}. Ultimately the goal is to estimate a concentration of dissolved carbon for a given observed pH value. Chemical equilibria, such as the carbonic acid system are inherently macroscopic definitions, often used in analytical chemistry in which solutions are commonly assumed to be completely homogeneous, having uniform concentrations throughout the entire solution. With that in mind, one can with relative ease deduce equilibrium concentrations and acidities of bodies of water in contact with carbon dioxide with a known partial pressure. However, in the system under consideration, the concentration fields are clearly not uniform, and thus the equilibrium models need to be applied differently.
Therefore a model is proposed, in which every small fluid element will be treated as being in pseudoequilibrium, such that the various forms of dissolved carbon are in equilibrium with each other, but not in equilibrium with the gaseous CO$_2$. This is intuitive in the sense that the fluid elements under consideration are below the gas-liquid interface. Thus each fluid element will be treated as having some total carbon content (which gives rise to its density increase) and some observable pH. Of the dissolved species, only the dissociated carbonic forms (\ch{H2CO3}, \ch{HCO3-}) affect pH, so for a given pH the concentration of dissociated carbonic forms can easily be found. If the underlying assumption that the dissolved species are in equilibria with each other is applied, this can in turn determine the total amount of dissolved carbon.
In order to use the equilibria to estimate a connection between pH and carbonic acid content, one needs to introduce two other useful equations, commonly used in analytical chemistry. These are the electrical charge conservation of the solution and the self-ionization equilibrium of water. It is noticeable that \ch{Na^+} was included in the calculations. That is because there is a fraction of \ch{NaOH} in the solution, which was used to neutralize the weak acidity of the pH indicators. The used chemicals and their amounts are showing in Tab. \ref{table:2}. 
The charge conservation requirement simply states that the charge weighted total concentration of positive and negative ions must cancel each other out, thus:  
\begin{equation}
    [H^+]+[Na^+]=[OH^-]+[HCO_3^-]+2[CO_3^{2-}]
\end{equation}
From the expression for $K_{a2}$ in Tab.\ref{table:1}, we find that: \\

\begin{equation}
[CO_3^{2-}]=[HCO_3^-]\frac{K_{a2}}{[H^+]}
\label{C2}
\end{equation}
Inserting this into the charge equation and applying the water self-ionization condition $[H^+][OH^-]=10^{-14}M^2$ we get: 
\begin{equation}
    [HCO_3^-]=\frac{1}{1+2K_{a2}/[H^+]}\left([Na]^++[H]^+-\frac{10^{-14}M^2}{[H^+]}\right)
    \label{C3}
\end{equation}
We then have an expression for the bicarbonate independent of the other carbon species. We now derive similar relations for the remaining two species, in terms of the bicarbonate
concentration from the expressions in Tab.\ref{table:1}: 
\begin{equation}
    [H_2CO_3]=[HCO_3^-]\frac{[H^+]}{K_{a1}}
\label{C4}
\end{equation}
\begin{equation}
    [CO_2]=[H_2CO_3]\frac{1}{K_H}=[HCO_3^-]\frac{[H^+]}{K_{a1}K_H}
\label{C5}
\end{equation}
By inserting  equations Eq.\ref{C2}, Eq.\ref{C4} and Eq.\ref{C5}  in the expression for total dissolved carbon:
\begin{equation}
   C_T=[CO_3^{2-}]+[HCO_3^-]+[H_2CO_3]+[CO_2]  
\end{equation}
we find:  \\
\begin{equation}
  C_T=[HCO_3^-] \left(\frac{K_{a2}}{[H^+]}+1+\frac{[H^+]}{K_{a1}}+\frac{[H^+]}{K_{a1}K_H}\right)
  \label{C7}
\end{equation}
Finally inserting Eq.\ref{C3} in Eq.\ref{C7} we get:
\begin{equation}
  C_T=\frac{1}{1+2K_{a2}/[H^+]}\left([Na]^++[H]^+-\frac{10^{-14}M^2}{[H^+]}\right)\left(\frac{K_{a2}}{[H^+]}+1+\frac{[H^+]}{K_{a1}}+\frac{[H^+]}{K_{a1}K_H}\right)
  \label{carb-con}
\end{equation}

where K$_H$, K$_{a1}$ and K$_{a2}$ are the equilibrium constants for the hydration of \ch{CO2} and dissociation steps of carbonic acid respectively.
Their values are given in Tab.\ref{table:1}. The expression above, along with 
the fact that \ch{[H^+]}$=10^{-\ch{pH}}$, gives an estimation of the total carbon
concentration of a given fluid element based on its observed pH.  
 
In general, equilibrium reactions are most commonly used in a macroscopic sense (i.e. to represent the entire cell). The presented pseudoequilibrium approach is only valid if the underlying equilibrium reactions have rates faster than the characteristic time scales of the convective transport. For non-equilibrium situations, one would have to solve the full reaction kinetic equations, which is not possible by only observing pH. Therefore this shortcoming is simply stated as fact, and all estimated carbon concentrations will assume that the characteristic time scales of the convective transport is significantly slower than those of the equilibrium reactions.

\begin{figure}
\centering
\includegraphics[width=1\linewidth]{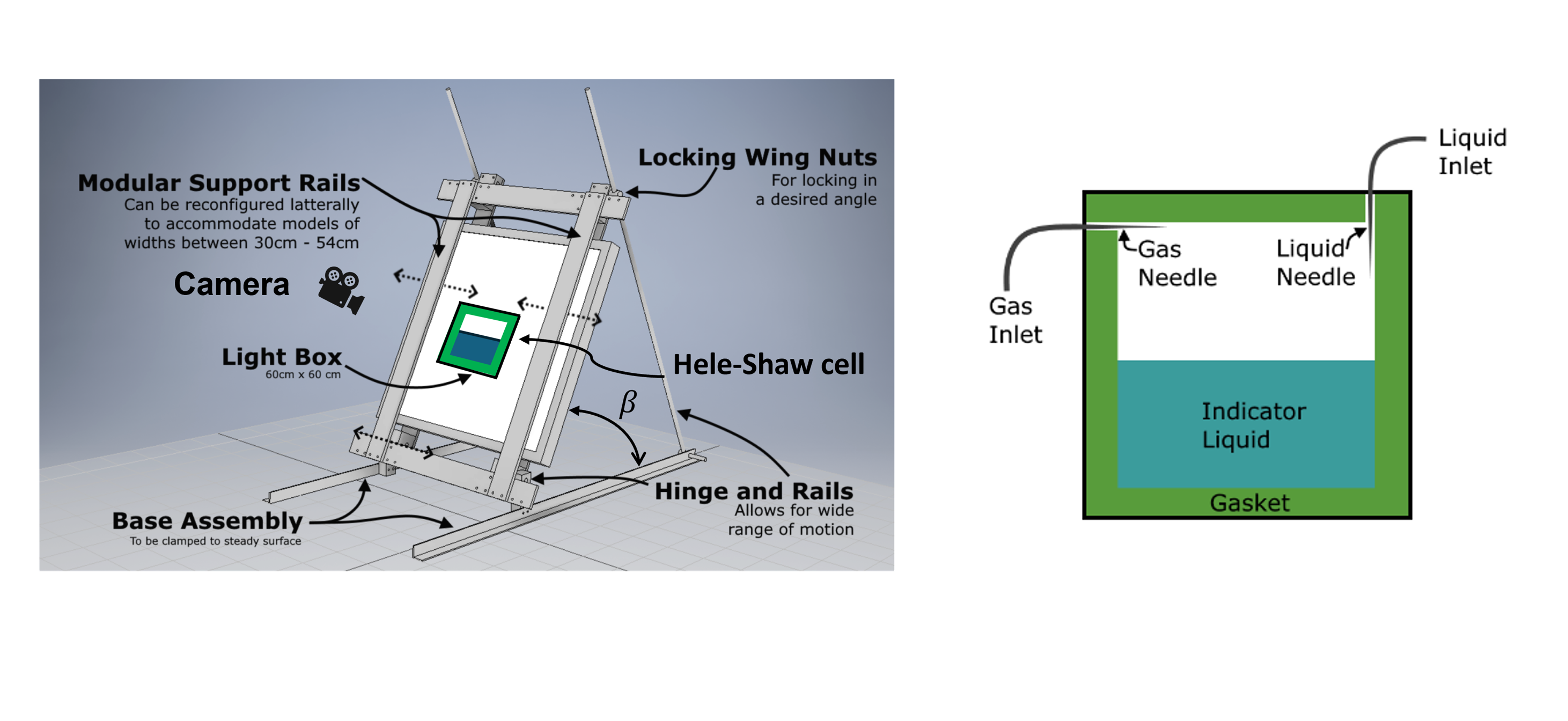}
\caption{Left: A CAD schematic of the experimental setup. A Hele-Shaw cell shown on the right was put on the LED light box, which  inclined angle $\beta$ can be adjusted by the mechanic frame. A high resolution camera was positioned in alignment with the flow cell and took images of the cell with regulated time interval during experiments. Right: A sketch of the Hele-Shaw cell with gasket. The liquid inlet needle is only used for filling up the cell, and the opening serves as a vent for excess gas during experiments. }
\label{Exp-model} 
\end{figure}

\begin{table}[h!]
\centering
\begin{tabular}{c |c | c | c} 
 \hline
  Equilibrium & Definition & Eq. constant & $pKa$ \\ [1.0ex] 
 \hline
 Gas dissolution & $K_h=\frac{[CO_2]}{p_{CO_2}}$ & $3.4 \cdot 10^{-2}$ M/atm & - \\
 Hydration  & $K_H=\frac{[H_2CO_3]}{[CO_2]} $ & $1.66 \cdot 10^{-3}$  & - \\
 First dissociation & $K_{a1}=\frac{[H^+][HCO_3^-]}{[H_2CO_3]}$ & $2.67 \cdot 10^{-4}$ M & $3.6$\\
 Second dissociation & $K_{a2}=\frac{[H^+][CO_3^{2-}]}{[HCO_3^-]}$ & $4.47 \cdot 10^{-11}$M  & $10.3$\\[1ex] 
 \hline
\end{tabular}
\caption{ Equilibrium constants relevant to the carbonic acid equilibrium at ambient conditions \cite{greenwood1997,sander2015} Acid $pKa$ values are presented for the dissociation equilibrium.}
\label{table:1}
\end{table}
\section{Experimental technique} \label{section3}
\subsection{The experimental model system}

A sketch of the model is shown on Fig.\ref{Exp-model}. The experiments are conducted in a Hele-Shaw cell, which is commonly used in the study of \ch{CO2} convective dissolution \cite{mojtaba2014,kneafsey2010,slim2013,wylock2011,thomas2015,loodts2014,backhaus2011,macminn2013,cardoso2014,outeda2014,tsai2013}. The Hele-Shaw cell  consists of two glass plates 35cm $\times$ 35cm with a thickness of 12mm and 16mm respectively. The glass plates are separated by a fixed distance, which in these experiments is 2.0mm.  A gasket (described below), with a flow domain of 32$\times$32cm is placed between the two glass plates.  Only the bottom half of this region is used, and two slits are made in the top, to allow for Luer lock syringe needles to be inserted into the cell.
The gasket is used to seal the model and to give a well defined distance between between the two glass plates. 
The needles are flattened to allow them to fit into the 2.0mm gap. 
 
 To illuminate the model we used a consumer grade LED light box model (IKEA FLOALT 60x60cm).  The experimental model is held in place by a mechanical frame made of machined aluminum blocks and 20x5mm profiles to which the light box is attached (See Fig.\ref{Exp-model}). This frame is hinged to a sturdy base at the bottom, and connected in a triangular fashion to a pair of cylindrical rails. This telescoping action along with locking wing nuts allows for setting the light box at any angle with respect to the ground  to be able to tune the component of the gravitational field along the model. The front-most profiles are connected to the rest of the assembly with M6 bolts, which can be attached in multiple places, such that experiments from 30cm to 54cm can be placed on the holder without visual obstruction from the rails.

  The gaskets are made of an addition-curing silicone mold-making compound (Koraform).  The benefits of this compound for creating a custom gasket are low pouring viscosity before hardening, fast curing, neglectable shrinkage and high mechanical strength. 
  After mixing the two components, a uniform slab is made by letting the silicone compound cure while being clamped between the two plates. 
  Metal spacers are inserted between the plates at the points of clamping to discourage warping of the plates, to achieve a thickness which is as uniform as possible.  After curing, slab is demolded and then trimmed to the desired dimensions.

\subsection{Visualizing acidity}
The raw data in the experiments conducted come in the form of images of the liquid under consideration. As previously mentioned, the carbonic acid equilibrium is a system of various dissolved inorganic carbon species which interact with each other, the solvent and the gas phase. 
Thus a relatively straight forward method of visualizing changes in acidity is to utilize pH indicators \cite{kneafsey2010,outeda2014,faisal2013,mojtaba2014}. This method is rather common in previously conducted work, but comes with some limitations \cite{thomas2015,dewit2016}.
\begin{figure}
\centering
\includegraphics[width=1\linewidth]{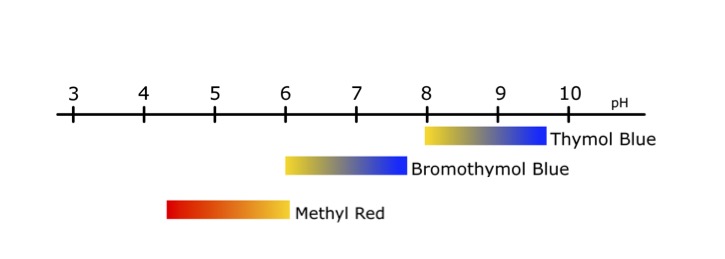}
\caption{A visual representation of the three pH indicators used. The colored regions show at which acidity values the indicators transition, and between what colors. Outside of the shown ranges the indicators hold a solid color. For instance Bromothymol Blue is active in a pH range from 6 to 7.6. In more acidic conditions it stays yellow, and in more basic conditions it stays blue.}
\label{Ph-indic} 
\end{figure}
A pH indicator is for all intents and purposes a weak acid, which has the added property that the protonated and dissociated forms of the molecule have different absorption spectra in the visible range. As per the definition of weak acids, this implies that each pH indicator has an associated equilibrium constant K$_a$, according to the reaction: \\ 
\begin{equation}
   HInd  \rightleftarrows H^+ + Ind^- \qquad
   K_{a}=\frac{[H^+][Ind]]}{[HInd]}
\end{equation}

This implies that there is a similar concentration of the protonated and dissociated forms of the indicator around pH$\simeq$pK$_a$.
Due to the logarithmic nature of the pH scale, the relative concentrations of these two forms grow and shrink exponentially around this acidity value. Therefore a color change is mostly observed when pH$\simeq$pK$_a$
Typically a deviation in pH of about 1 logarithmic unit makes either form completely dominate, and little color change is observed past this point. For the application in question, this means that any single pH indicator is only useful to visualize a limited range of carbonic acid concentration. Therefore, the choice of pH indicator inherently sets an upper and lower bound on the concentrations one can deduce.

To combat this limitation, three different pH indicators are combined in an attempt to form a more continuous color spectrum (see Tab.\ref{table:2}), such that more information about the pH fluctuations can be extracted from the image data. Fig.\ref{Ph-indic} shows the active ranges and colors of the pH indicators considered for this purpose. The goal is to produce a solution of these indicators that is as active as possible down to the equilibrium pH of water in contact with atmospheric pressure of pure \ch{CO2}. Thus the ideal solution goes through significant and distinctive color variations from neutral conditions, to a pH of about 4.
\begin{table}[h!]
\centering
\begin{tabular}{c |c | c} 
 \hline
  & Stock & 1:50 Dilution \\ [1.0ex] 
 \hline
 Thymol blue [mg/L] & 1250 & 25 \\
 Bromothymol blue [mg/L] & 2500 & 50 \\
 Methyl red [mg/L] & 1600 & 32 \\
\hline
 2-propanol \%vol & 12.5 & 0.25\\
 NaOH [mol/L] & $2.5 \cdot 10^{-3}$ & $5.0 \cdot 10^{-5}$\\[1ex] 
 \hline
\end{tabular}
\caption{ The wide-range pH indicator developed for \ch{CO2} DDC experiments. The first column is a stock solution, and the second column is the 1:50 dilution of the stock, which is used for experiments. In addition to the color content, the solvent additives are shown below. }
\label{table:2}
\end{table}

As previously explained, the fundamental working principle of pH indicators is that they themselves are weak acids. 
This by definition implies that their respective acid-base equilibria couple with the carbonic acid equilibrium system under consideration. 
Upon taking molar weights into account, one can check from Tab.\ref{table:2} that the concentration of the color components are of the order of magnitude $10^{-5}$M. 
Literature suggests that the total concentration of CO$_2$ derivatives in water in equilibrium with atmospheric pressure of pure CO$_2$ is of the order of 
magnitude $10^{-2}$M, which is three order of magnitude greater than that of the indicators \cite{greenwood1997}.
That being said, the concentration of the individual carbonic species can be significantly lower.  Thus the use of color indicators in general could in theory interfere with the very thing under observation. This potential issue is simply postulated here, and the later analysis will neglect any chemical equilibrium interference which might arise from these interference effects. Previous literature suggests this should be a valid assumption \cite{thomas2015}.

\begin{figure}
\centering
\includegraphics[width=0.7\linewidth]{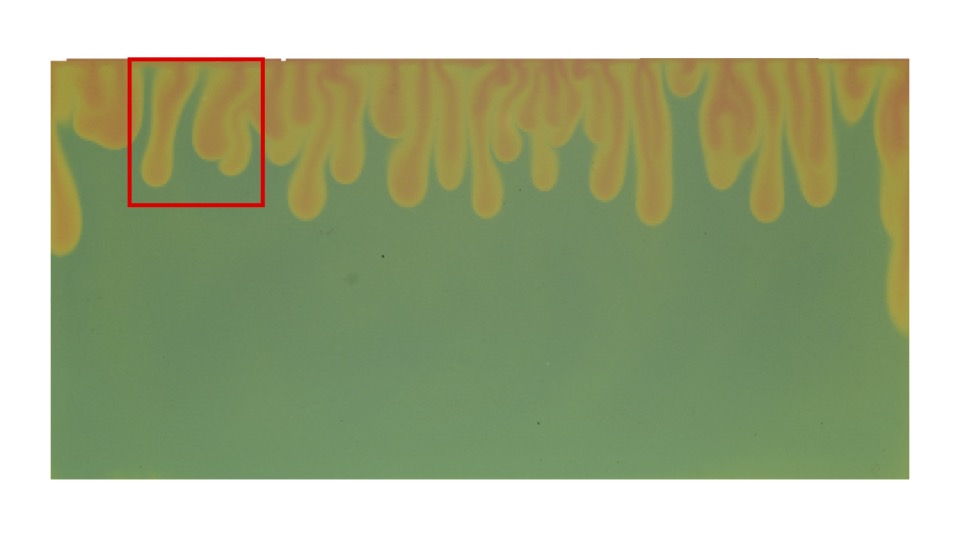}
\caption{Image after perspective transformation and masking procedure. This corresponds to  $25$cm $\times$ $12$cm in the experimental model system. The red rectangle shows the cropped part used in Fig.\ref{pH-pic} a). }
\label{Conv-pic} 
\end{figure}

\subsection{Image analysis}
The applied image analysis consists of taking the raw image, applying a simple geometrical transformation to it, identifying and masking the liquid region and determining the pH field from the color of the liquid.
\begin{figure}
\centering
\includegraphics[width=1\linewidth]{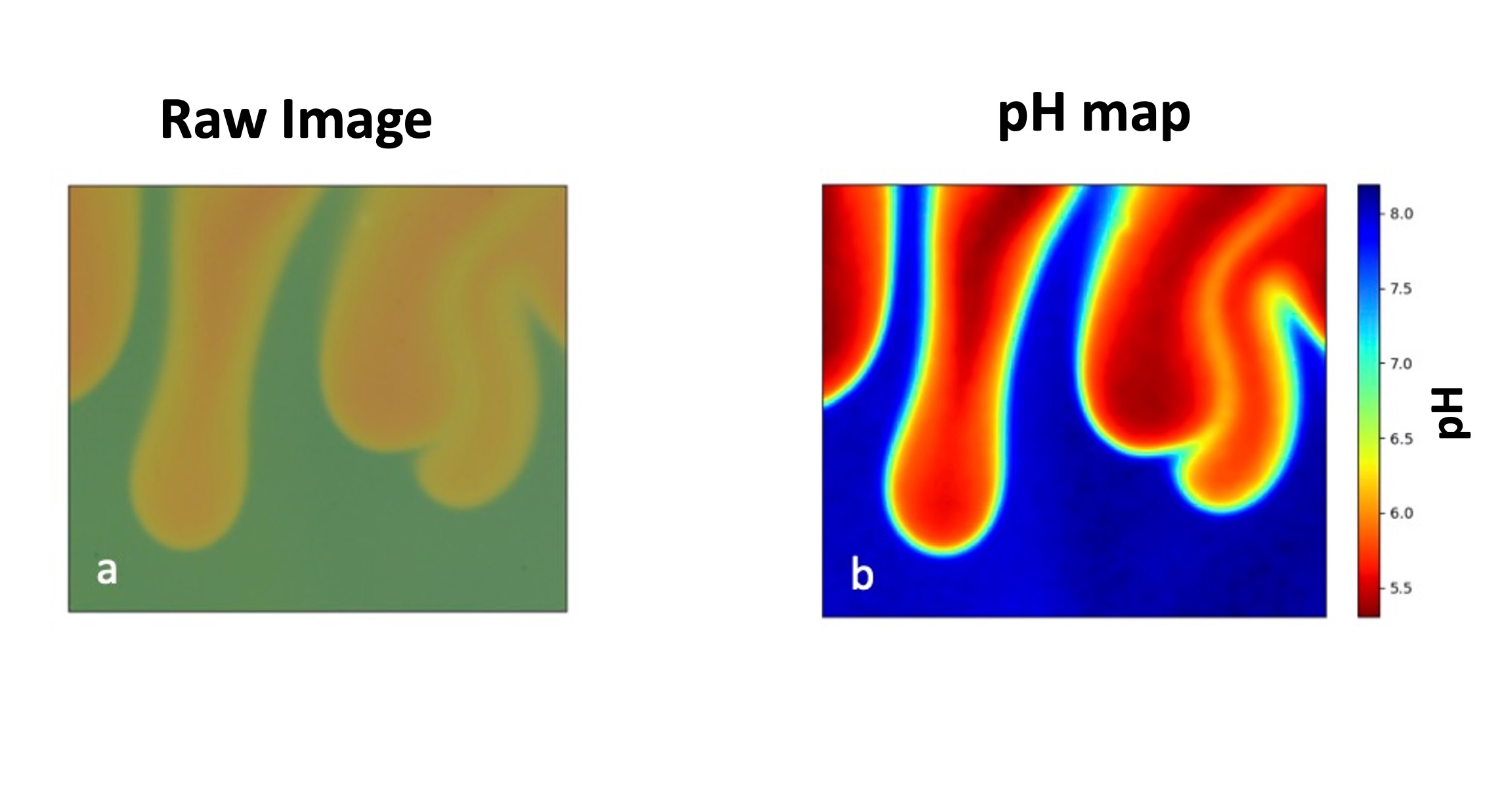}
\caption{An example input and output of the complete pH determination method. With this pH indicator mixture and color analysis method, a continuous pH field is determined, yielding much more information than a conventional thresholding and/or edge detection method. a) A cropped region of the preprocessed frame Fig.\ref{Conv-pic}.  b) The result of the pH determination algorithm where the color codes give the pH values in the image.}
\label{pH-pic} 
\end{figure}
Before applying the color interpolating method in Section \ref{interpolation}, every picture frame needs to be pre-processed geometrically and masked to display only the liquid region. Since the camera is never perfectly aligned with respect to the flow cell, a bounding box for the liquid region is found and a perspective transformation is applied to transform this into a rectangle. By applying this perspective transformation, the plane of the experiment now matches the plane of the transformed image. Afterwards, a mask is applied to identify the non-liquid regions. The masked pixels are then removed, and do not take part in the color interpolation algorithm. An example of a final picture after perspective transformation and masking procedure is shown in Fig.\ref{Conv-pic}.
\subsection{Interpolating colors} \label{interpolation}
Now that the fluid region has been isolated, the task at hand is finding a meaningful way of correlating a color to a pH level. From the raw image file, the color of each pixel is represented by three 8-bit integers, for respectively the red, green and blue channel. For all intents and purposes, this can be thought of as a 3D vector space, and each pixel can be treated as an element of said vector space.

By interpreting the three channels as coordinates of a vector space, we can produce a scatter plot that visualizes the color change undergone by the liquid throughout the experiment. Consider a region Fig.\ref{pH-pic} a) cropped from the center of the final frame of Fig.\ref{Conv-pic} (red square in figure). The region is chosen to be representative of the color spectrum the interpolation method must deal with. 

\begin{figure}[H]
\centering
\includegraphics[width=0.9\linewidth]{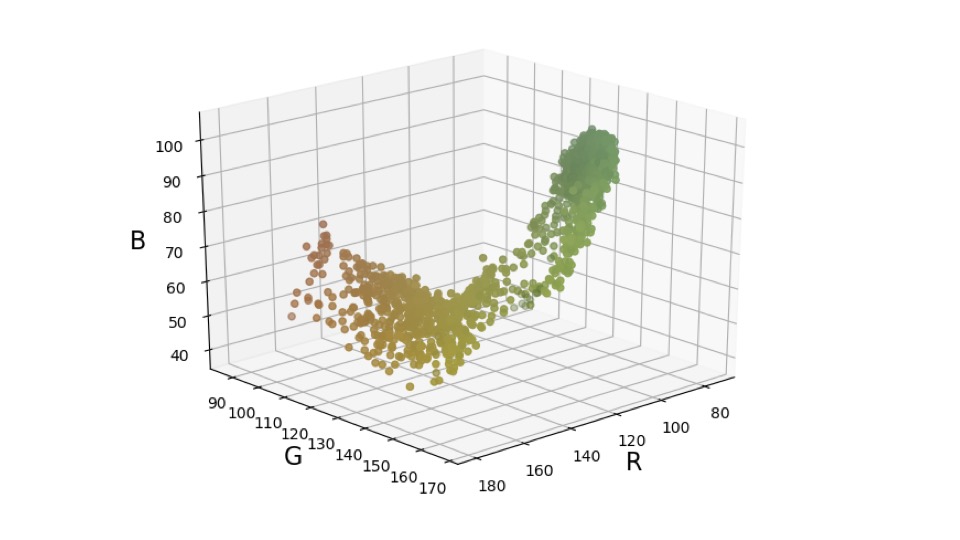}
\caption{The unaltered RGB space where each dot represents the color of a pixel in the original raw RGB image. The acidification process can be viewed as a transformation in this space, where each dot moves as its color varies.}
\label{RGB-plot} 
\end{figure}

\begin{figure}[H]
\centering
\includegraphics[width=1\linewidth]{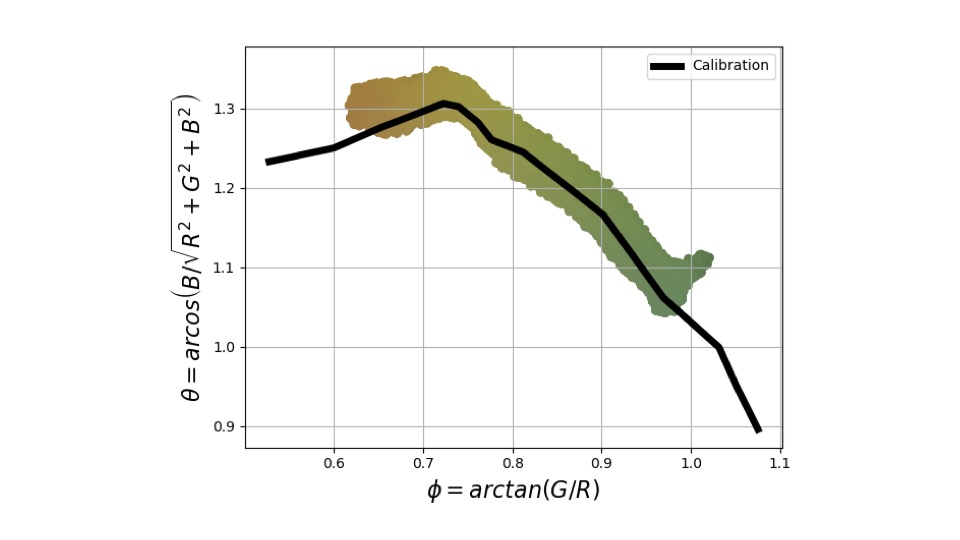}
\caption{The normalized RGB space represented by a polar and azimuthal angle. The black path is calibrated with known pH values.}
\label{RGB-polar} 
\end{figure}

By interpreting the three channels as coordinates of a vector space, we can also produce a scatter plot that shows the placement of each individual pixel in the color space. Upon inspection of Fig.\ref{RGB-plot}, one can see that representative collection of pixels are scattered around a path in the three dimensional color space.
If the coordinates of the color space are transformed from typical Cartesian (R,G,B) to spherical, one obtains a radial coordinate (analogous to brightness) and two angular coordinates, which represent the colors of pixels as shown in Fig.\ref{RGB-polar}. Since the brightness of a pixel has no effect on which pH value should be assigned to it, this component can be discarded. The valuable information is in the polar and azimuthal angles, which can then be used to correlate a pixel color to a pH value.


\begin{figure}[H]
\centering
\includegraphics[width=0.9\linewidth]{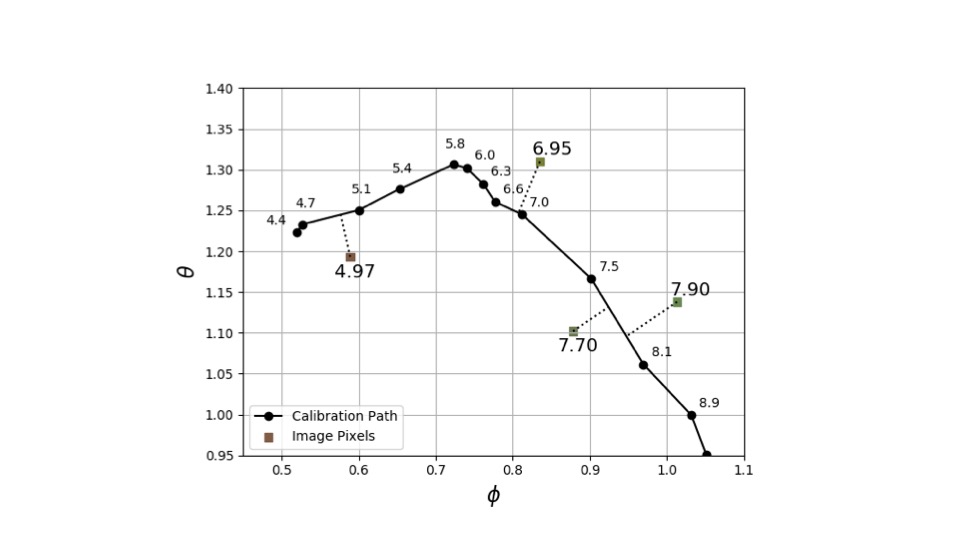}
\caption{A visual representation of the pH determination procedure. Here a pH value is determined for 4 example colors, by finding their closest point on the calibration curve.}
\label{Polar-closest} 
\end{figure}

By titrating the pH indicator solution to
various known acidity levels, one can construct a calibration path by injecting the
mixtures with known pH levels into the flow
cells and imaging them. For each calibration image, a uniform region, free of air bubbles and other artifacts is found, and used
to produce an averaged color value associated with that pH value. This is repeated multiple times, to produce a 
calibration curve Fig.\ref{Polar-closest}. To deduce the pH of any
unknown pixel, one then finds the shortest
distance from it to this calibrated path in
the angular representation of the RGB color
space. The shortest distance from any point
to the calibration path can be found analytically (See Fig.\ref{Polar-closest}), and said point is then assigned to the
line segment to which the distance is shortest. On this line segment, the distance between the two neighbouring calibration colors can be used to interpolate the pH value, which will then be assigned to the pixel in question. This method is less sensitive to image noise, as it relies on both coordinates, rather than for instance interpolating pH as a function of one value. Despite this, a small Gaussian filter is still applied to the resulting pH field, to filter out pixel-scale fluctuations, and to combat the fact that the raw image data is discrete. The end result of the pH determination algorithm is shown in Fig.\ref{pH-pic} b), where the input image is the cropped section Fig.\ref{pH-pic} a) of Fig.\ref{Conv-pic}.

\section{MEASUREMENT OF \MakeLowercase{p}H AND  CARBON CONCENTRATION} \label{section4}

\begin{figure}
\centering
    \includegraphics[width=1.0\linewidth]{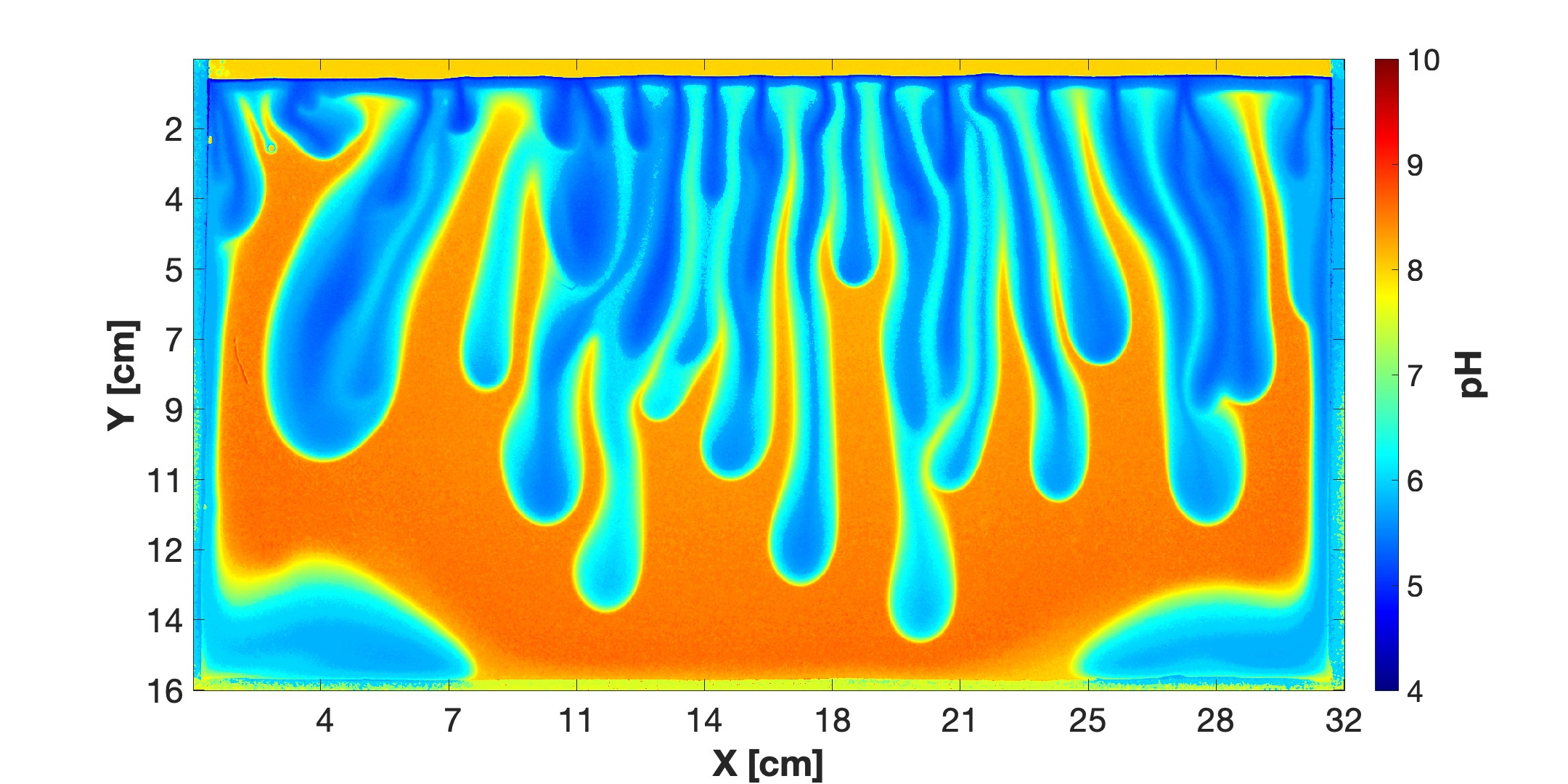}
    \caption{The pH field with a tilt angle of 60° and Hele-Shaw dimensions of 32cm$\times$16cm$\times$2mm, 46 min after \ch{CO2} injection. The plumes appear to shift slightly laterally to the edges. The color shows spatial pH values. The link of the video showing the overall evolution of pH filed is attached in the supplementary material \cite{spatiotemporal_videos}.}
    \label{pH_60}
\end{figure}

\begin{figure*}
\centering
    \includegraphics[width=1.0\textwidth]{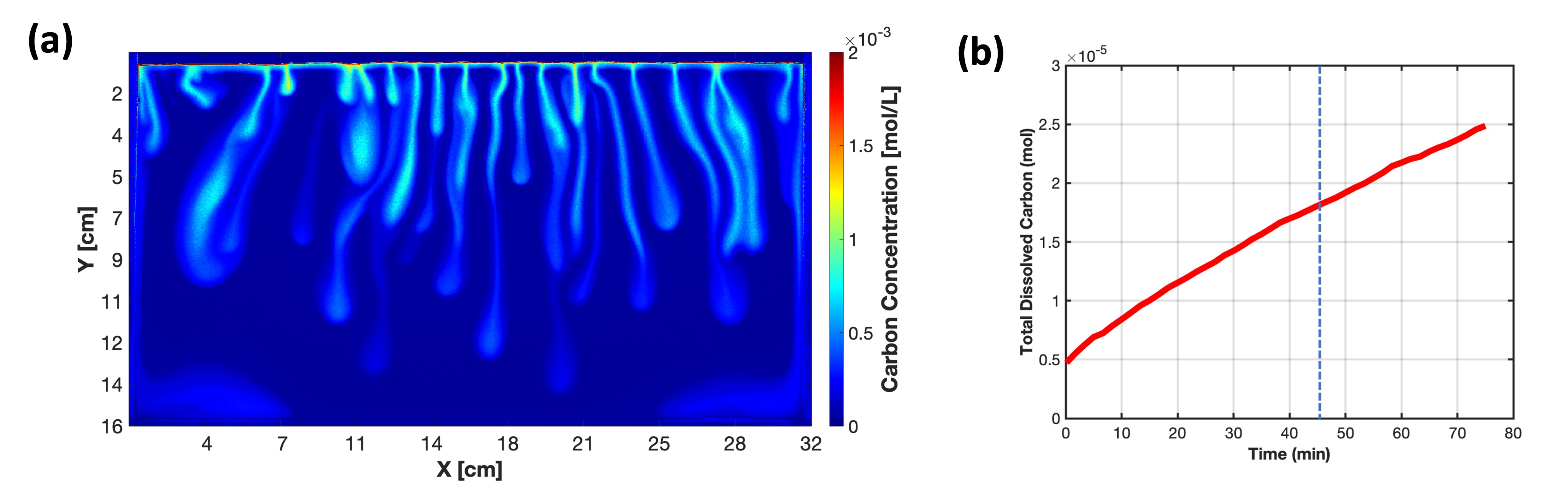}
    \caption{a) The carbon concentration field with a tilt angle of 60° and Hele-Shaw dimensions of 32cm$\times$16cm$\times$2mm, 46 min after \ch{CO2} injection. The color shows spatial carbon concentration values. The plume patterns correspond to the pH pattern in Fig.\ref{pH_60}. As the concentration of dissolved carbon increases, the pH level decreases. b) The evolution of total dissolved carbon in the whole Hele-Shaw cell. The experiment was conducted with a tilt angle of 60° and Hele-Shaw dimensions of 32cm$\times$16cm$\times$2mm. The blue dash line in b) corresponds the time in a). It showed there were around 1.8$\cdot$10$^{-5}$ mol carbon dissolved in a). The link of the video showing the overall evolution of carbon concentration is attached in the supplementary material \cite{spatiotemporal_videos}.}
    \label{Carbon and total carbon}
\end{figure*}

Images obtained of flow patterns in the refined 2mm cell were used along with the tailored pH color analysis method and Eq.\ref{carb-con} to determine the pH and concentration fields. The convection plumes for a tilt angle of $\theta$ = 60° and Hele-Shaw dimensions of 32cm$\times$16cm$\times$2mm are shown in Fig.\ref{pH_60} and Fig.\ref{Carbon and total carbon}. The results clearly indicate that there is a tendency for asymmetric flow. We see that the plumes move towards the edges and are longer in the central part of the model. The videos displaying the overall migration and morphological evolution of plumes can also be visualized in the supplementary material \cite{spatiotemporal_videos}. This is most likely due to permeability fluctuations. The clamping pressure could be warping the plates slightly, which could be enough to cause minor fluctuations in effective plate spacing.

Despite the transverse deviation of the plumes, the obtained concentration field data does demonstrate that the developed pH indicator method can indeed account for concentration variations within and around convection plumes and give an estimate of the pH field, concentration field, and total amount of dissolved carbon as shown in Fig.\ref{pH_60}, Fig.\ref{Carbon and total carbon} a) and Fig.\ref{Carbon and total carbon} b) respectively. These are data  which can not be well defined with single indicator methods or Schlieren imaging. Despite the asymmetrical flow, the results do demonstrate how the indicator, color analysis, and chemical equilibrium model can help investigate the continuous nature of the spatiotemporal pH and carbon concentration.

\section{Conclusion}
The inspiration for this project came mainly from previous work on plume patterns in DDC
\cite{szulczewski2012,dewit2016,thomas2015,riaz2006,backhaus2011,macminn2013,wooding1969,faisal2013,faisal2015}, and especially an article by Thomas et al. \cite{thomas2015} which demonstrated the importance of proper use of pH color indicators for visualization of the acidic plume formations. This was an inspiration to attempt to derive a more rigorous method by using the principles of analytical chemistry, a tailored color indicator mixture and image analysis to shed light on the continuous nature of the pH and concentration fields that are often neglected in similar experiments. Most comparable experiments rely on a single indicator only, and thus put a limited range one can detect, beyond which most concentration values remain undetectable. However, the techniques explored in this manuscript for the quantitative measurement of carbon concentrations can still find potential applications in several other setups where convective dissolution is explored \cite{brouzet2022co,taheri2021effect,tang2019experimental}.

While the experimentally obtained plume morphologies are heavily affected by experimental artifacts like non uniform plate separation, the developed color analysis method was clearly able to identify the varying pH field, and coupled with the assumed pseudoequilibrium model, was able to estimate the carbon concentration fields. Therefore one can conclude that using a universal pH indicator, or a mixture of multiple individual color indicators can in fact prove useful to visually investigate patterns of fluids in which chemical reactions take place. A recommendation for further study within this topic would be to more rigorously validate at which indicator-concentrations does the presence of an indicator significantly alter the system dynamics. For this system it has been argued that the concentrations used do not affect the system in a meaningful way \cite{thomas2015}, but that may not be the case for other systems one might want to apply a similar approach to.

\section*{Acknowledgement}

We acknowledge the financial support from the Research Council of Norway through the PoreLab Center of Excellence (project number 262644) and the FlowConn Researcher Project for Young Talent (project number 324555). We also thank the support from the Njord Center, Faculty of Mathematics and Natural Sciences at the University of Oslo through the project CO2Basalt.


\newpage
\bibliographystyle{Frontiers-Vancouver} 
\bibliography{biblio}
\end{document}